\documentclass[letterpaper,english,reprint, aps]{revtex4-1}
\usepackage[T1]{fontenc}
\usepackage[latin9]{inputenc}
\usepackage{amsmath}
\usepackage{amssymb}

\makeatletter

\pdfpageheight\paperheight
\pdfpagewidth\paperwidth

\makeatother

\usepackage{babel}
\begin{document}
\title{Pfaffian formulas for non equivalent bases}
\author{L.M. Robledo}
\email{luis.robledo@uam.es}

\affiliation{Departamento de Física Teórica and CIAF, Universidad Autónoma de Madrid,
E-28049 Madrid, Spain}
\affiliation{Center for Computational Simulation, Universidad Polit\textbackslash 'ecnica
de Madrid, Campus de Montegancedo, Boadilla del Monte, E-28660-Madrid,
Spain}
\date{\today}
\begin{abstract}
Pfaffian formulas used to compute overlaps necessary to carry out
generator coordinate method calculations using a set of Hartree- Fock-
Bogoliubov wave functions, is generalized to the case where each of
the HFB states are expanded in different arbitrary bases spanning
different sub-space of the Hilbert space. The formula obtained is
compared with previous results proving to be completely equivalent
to them. A discussion of equivalent formulas obtained in the literature
is carried out.
\end{abstract}
\maketitle

\section{\protect\label{sec:Intro}Introduction}

The evaluation of operator overlaps for arbitrary mean field wave
functions of the Hartree- Fock (HF or Slater) or Hartree- Fock- Bogoliubov
(HFB) type is of great interest for the many applications in the subject
of symmetry restoration and generator coordinate (GCM) or configuration
interaction like (CI) methods (see recent reviews in Refs \citep{Sheikh2021}
and \citep{Piela2020}. It turns out that the formulas developed earlier
in the literature for the HFB case are not valid when the HFB states
(or their transformed under symmetry operations) are expanded in bases
which are non-unitarily equivalent. To overcome this difficulty one
usually invokes the formal extension of the original bases as to make
them complete by adding states having zero occupancy. This approach
was pursued in Refs \citep{BONCHE1990466,Valor2000145} for unitary
and in Ref \citep{PhysRevC.50.2874} for general canonical transformations.
Recently, the formalism of Ref \citep{PhysRevC.50.2874} has been
extended \citep{Robledo2022a} as to give formulas which manifestly
depend only on quantities defined in the original bases and therefore
they are completely independent of the added basis vectors. In a subsequent
paper, the formalism was applied to the common situation when the
HFB states are expanded in harmonic oscillator (HO) basis with different
oscillator lengths \citep{Robledo2022}. In this work, it was clearly
demonstrated that the issue with non-equivalent bases cannot be overlooked
as its consideration leads to substantial differences in the computed
overlap between HFB states. These ideas are also taken into account
in the implementation of angular momentum projection by using full
HO major shells in order to use the traditional formulas. In the context
of angular momentum projection the formalism of \citep{PhysRevC.50.2874}
has recently been used in \citep{Marevic2020} to consistently compute
the rotational correction to the potential energy surfaces of fission.
Symmetry restoration in a spatial domain has received lately a lot
of attention in connection with the proper definition of quantum numbers
in fission fragments. For instance, particle number restoration in
fission fragments has been discussed by Simenel \citep{Simenel2010}.
As discussed elsewhere \citep{robledo2025}, symmetry restoration
in a domain fully fall in the category of non-equivalent bases as
the operators used to limit the domain are only defined in the whole
Hilbert space. On the other hand, the formalism for non-equivalent
basis was incorporated into the pfaffian formalism \citep{PhysRevC.79.021302}
first in Ref \citep{PhysRevC.84.014307} by the present author and
subsequently in Ref \citep{Avez2012} by Avez et al. In the first
paper, instead of using a formal extension to an infinite basis as
in Ref \citep{PhysRevC.50.2874} I used the union of the two bases
after proper orthogonalization. The formula obtained is rather involved
specially after a comparison with the one given in Ref \citep{Avez2012}.
However, as discussed below the derivation of the result of \citep{Avez2012}
concerning non-equivalent bases contains some inconsistencies that
need attention, in spite of the fact that the formula in \citep{Avez2012}
and the one obtained here are equivalent. Also in the formula obtained
in \citep{Avez2012} the inverse of the overlap matrix appears, an
inconvenient feature that is not present in the result discussed here.
The purpose of this paper is to obtain a pfaffian formula for the
overlap in the case of non-equivalent bases by using the formal extension
of the bases to infinite ones. A comparison with other results used
in the literature is made and some potentially dangerous situations
are pointed out. Finally, it is proven how the result of \citep{PhysRevC.50.2874,Robledo2022a}
is fully recovered. 

\section{Pfaffian formula for non-equivalent bases}

As in the derivation of Refs \citep{PhysRevC.50.2874} the finite
basis system is portrayed as an infinite Hilbert space with Bogoliubov
amplitudes $U_{i}$ and $V_{i}$ 
\begin{equation}
V_{i}=\left(\begin{array}{cc}
\bar{V}_{i} & 0\\
0 & 0
\end{array}\right),\;\;U_{i}=\left(\begin{array}{cc}
\bar{U}_{i} & 0\\
0 & d_{i}
\end{array}\right),\label{eq:block-1}
\end{equation}
where $\bar{V}_{i}$and $\bar{U}_{i}$ are the $N_{i}\times N_{i}$
matrices in the finite bases, $\mathcal{B}_{i}=\{c_{i,k}^{\dagger},k=1,\ldots,N_{i}\}$.
From now on we assume $N_{0}=N_{1}=N=2n$ which is not a serious limitation
as one can choose $N$ as the largest of $N_{i}$ and trivially enlarge
the Bogoliubov amplitudes of the other system. The dimensionality
of $V_{i}$ and $U_{i}$ is infinite and corresponds to an expansion
in the bases $\mathcal{B}_{i}\cup\bar{\mathcal{B}}_{i}=\{c_{0,k}^{\dagger}\}^{\infty}$
where $\bar{\mathcal{B}}_{i}=\{c_{i,k}^{\dagger},k=N+1,\ldots,\infty\}$
is the complement of $\mathcal{B}_{i}$ in the whole Hilbert space.
A unitary matrix $d_{i}$is introduced in the complementary space.
The result will also prove to be independent of $d_{i}$. One also
needs the overlap matrix $R_{kl}=\{c_{0,k}^{\dagger},c_{1,l}\}=_{0}\langle k|l\rangle_{1}$
($k,l=1,\ldots,\infty$) between the two complete bases $\mathcal{B}_{0}\cup\bar{\mathcal{B}}_{0}$
and $\mathcal{B}_{1}\cup\bar{\mathcal{B}}_{1}$ as well as its block
decomposition
\begin{equation}
R=\left(\begin{array}{cc}
\mathcal{R} & \mathcal{S}\\
\mathcal{T} & \mathcal{U}
\end{array}\right)\label{eq:Rblock}
\end{equation}
in terms of the restricted overlap $\mathcal{R}_{kl}$ ($k,l=1,\ldots,N$),
and the remaining blocks. To facilitate the discussion an integer
$M$ which is allowed to tend to infinity at the end is introduced
as the dimension of the Hilbert space. As shown below the final result
do not depend on this quantity and therefore the limit $M\rightarrow\infty$
can be safely taken. With these definitions we can use the pfaffian
formula in the $M$ dimensional space
\begin{equation}
\langle\phi_{0}|\phi_{1}\rangle=s_{M}\textrm{pf \ensuremath{\left(\begin{array}{cc}
RM^{(1)}R^{T} & -\mathbb{I}\\
\mathbb{I} & -M^{(0)\,*}
\end{array}\right)}}=s_{M}\textrm{pf}\mathbb{M}\label{eq:PfaffOri}
\end{equation}
with $M^{(i)}=\left(V_{i}U_{i}^{-1}\right)^{*}$ and the relation
$c_{1,l}^{+}=\sum_{k}R_{kl}^{*}c_{0,k}^{+}$ with the unitary transformation
$R$ has been used. Please note that $R$ is connecting two complete
bases and therefore is unitary, a property that do not hold for $\mathcal{R}$.
In order to show how the above expression reduces to one where the
quantities refer to the finite bases only, we will make use of the
block structure of the matrix that appears as the argument of the
pfaffian and its properties under the exchange of rows and columns.
First, it is straightforward to obtain
\begin{equation}
M^{(0)}=\left(\begin{array}{cc}
\bar{M}^{(0)} & 0\\
0 & 0
\end{array}\right)\label{eq:Mi}
\end{equation}
where the matrix $\bar{M}^{(i)}=\left(\bar{V}_{i}\bar{U}_{i}^{-1}\right)^{*}$
of dimension $N\times N$ are introduced. In the next step the product
$RM^{(1)}R^{T}$ is expanded as
\begin{align*}
RM^{(1)}R^{T} & =\left(\begin{array}{cc}
\mathcal{R} & \mathcal{S}\\
\mathcal{T} & \mathcal{U}
\end{array}\right)\left(\begin{array}{cc}
\bar{M}^{(1)}\mathcal{R}^{T} & \bar{M}^{(1)}\mathcal{T}^{T}\\
0 & 0
\end{array}\right)\\
 & =\left(\begin{array}{cc}
\mathcal{R}\bar{M}^{(1)}\mathcal{R}^{T} & \mathcal{X}_{12}\\
\mathcal{X}_{21} & \mathcal{X}_{22}
\end{array}\right)
\end{align*}
where $\mathcal{X}_{12}=\mathcal{R}\bar{M}^{(1)}\mathcal{T}^{T}$
is a $N\times(M-N)$ matrix. The structure of the other two $\mathcal{X}_{21}=-\mathcal{X}_{12}^{T}$
and $\mathcal{X}_{22}=\mathcal{T}\bar{M}^{(1)}\mathcal{T}^{T}$(skew-symmetric)
can be easily obtained, but both matrices as well as $\mathcal{X}_{12}$
are irrelevant for the final result. Applying now the ``move and
shift'' operation $S(i,j)$ (see appendix A of Ref \citep{PhysRevC.84.014307})
to the $N$ columns of $\mathbb{M}$starting at column $M+1$ to bring
them to column $N+1$ (and the same for the corresponding rows) one
obtains
\[
\textrm{pf}\mathbb{M}=f\begin{pmatrix}\mathcal{R}\bar{M}^{(1)}\mathcal{R}^{T} & -\mathbb{I}_{11} & \mathcal{X}_{12} & 0\\
\mathbb{I}_{11} & -M^{(0)\,*} & 0 & 0\\
\hline \mathcal{X}_{21} & 0 & \mathcal{X}_{22} & -\mathbb{I}_{22}\\
0 & 0 & \mathbb{I}_{22} & 0
\end{pmatrix}
\]
where we have introduced the identity matrices $\mathbb{I}_{11}$
and $\mathbb{I}_{22}$ of dimension $N\times N$ and $(M-N)\times(M-N)$,
respectively and the phase $f=(-1)^{(M-N)N}$. We are now in the position
to use the formula for the pfaffian of a block matrix (see appendix
B of \citep{PhysRevC.84.014307})
\[
\textrm{pf}\mathbb{M}=f\textrm{pf}\left(\begin{array}{cc}
\mathcal{R}\bar{M}^{(1)}\mathcal{R}^{T} & -\mathbb{I}_{11}\\
\mathbb{I}_{11} & -M^{(0)\,*}
\end{array}\right)\textrm{pf}\left(\begin{array}{cc}
\mathcal{Y}_{22} & -\mathbb{I}_{22}\\
\mathbb{I}_{22} & 0
\end{array}\right)
\]
where the skew-symmetric matrix $\mathcal{Y}_{22}$ has dimension
$(M-N)\times(M-N)$. Its explicit form is irrelevant as one of the
properties of the pfaffian tell us 
\[
\textrm{pf}\left(\begin{array}{cc}
\mathcal{Y}_{22} & -\mathbb{I}_{22}\\
\mathbb{I}_{22} & 0
\end{array}\right)=(-1)^{(M-N)(M-N+1)/2}
\]
After collecting all the phases one arrives to 
\begin{equation}
\langle\phi_{0}|\phi_{1}\rangle=s_{N}\textrm{pf \ensuremath{\left(\begin{array}{cc}
\mathcal{R}\bar{M}^{(1)}\mathcal{R}^{T} & -\mathbb{I}\\
\mathbb{I} & -\bar{M}^{(0)\,*}
\end{array}\right)}}=s_{N}\textrm{pf}\bar{\mathbb{M}}\label{eq:PfaffFin}
\end{equation}
 which is the final expression. The overlap is given in terms of quantities
defined in the original bases and therefore the dimension of the argument
of the pfaffian is $(2N)\times(2N).$ By comparing this result with
the one of Eq \eqref{eq:PfaffOri} one observes many similarities,
but there are subtle and important differences. The matrix in Eq \eqref{eq:PfaffOri}
is a $(2M)\times(2M)$ matrix, the overlap $R$ is a $M\times M$
unitary matrix connecting the bases $\mathcal{B}_{0}\cup\bar{\mathcal{B}}_{0}$
and $\mathcal{B}_{1}\cup\bar{\mathcal{B}}_{1}$ and the matrices $M^{(i)}$
are the ones of Eq \eqref{eq:Mi}. The result generalizes the one
of Eq (7) of \citep{PhysRevC.79.021302} and represents the main finding
of the paper. Please note that along the derivation there is no explicit
need to consider the inverse of the matrix $\mathcal{R}$which represents
a simplification with respect to other formulas (see below).

It is important now to connect the above result with the one of \citep{PhysRevC.50.2874}.
For this purpose one can use Eq (8) of \citep{PhysRevC.79.021302}
to write, up to a sign,
\[
\langle\phi_{0}|\phi_{1}\rangle=\left(\det\left(\mathbb{I}+\bar{M}^{(0)\,+}\mathcal{R}\bar{M}^{(1)}\mathcal{R}^{T}\right)\right)^{1/2}
\]
The argument of the determinant can be written as $\left(U_{0}^{-1}\right)^{T}\left(U_{0}^{T}(\mathcal{R}^{T})^{-1}U_{1}^{*}+V_{0}^{T}\mathcal{R}V_{1}^{*}\right)\left(U_{1}^{*}\right)^{-1}\mathcal{R}^{T}$
and therefore
\begin{align*}
\langle\phi_{0}|\phi_{1}\rangle & =\left(\det U_{0}\det U_{1}^{*}\right)^{-1/2}\times\\
\times & \left(\det\left(U_{0}^{T}(\mathcal{R}^{T})^{-1}U_{1}^{*}+V_{0}^{T}\mathcal{R}V_{1}^{*}\right)\det\mathcal{R}\right)^{1/2}
\end{align*}
which is Eq (25) of \citep{Robledo2022a} with $A=U_{0}^{T}(\mathcal{R}^{T})^{-1}U_{1}^{*}+V_{0}^{T}\mathcal{R}V_{1}^{*}$.
The extra factor $\left(\det U_{0}\det U_{1}^{*}\right)^{-1/2}$ takes
into account the different normalization of the HFB wave functions
in this paper and in Ref \citep{Robledo2022a}. 

\section{Comparison with other approaches}

Another equivalent expression to the one in Eq \eqref{eq:PfaffFin}
can be obtained by using the decomposition of 
\begin{align*}
\mathbb{M}=\left(\begin{array}{cc}
\mathcal{R}\bar{M}^{(1)}\mathcal{R}^{T} & -\mathbb{I}\\
\mathbb{I} & -\bar{M}^{(0)\,*}
\end{array}\right)
\end{align*}
as
\[
\mathbb{M}=\left(\begin{array}{cc}
\mathcal{R} & 0\\
0 & \mathbb{I}
\end{array}\right)\left(\begin{array}{cc}
\bar{M}^{(1)} & -\mathcal{R}^{-1}\\
\left(\mathcal{R}^{T}\right)^{-1} & -\bar{M}^{(0)\,*}
\end{array}\right)\left(\begin{array}{cc}
\mathcal{R}^{T} & 0\\
0 & \mathbb{I}
\end{array}\right)
\]
that leads to 
\[
\langle\phi_{0}|\phi_{1}\rangle=s_{N}\det\mathcal{R}\textrm{pf \ensuremath{\left(\begin{array}{cc}
\bar{M}^{(1)} & -\mathcal{R}^{-1}\\
\left(\mathcal{R}^{T}\right)^{-1} & -\bar{M}^{(0)\,*}
\end{array}\right)}}
\]
which is Eqs (61) and (54) of \citep{Avez2012}. In this paper, the
authors consider the possibility of having two non-unitary equivalent
basis in their derivation but they do not consider explicitly the
extension to an (infinite) complete basis like it is done here. As
a consequence, the introduction of the product of two pseudo-identities
\footnote{They are called pseudo-identities as their definition in terms of
Grassman numbers in Eq (19) \citep{Avez2012} only consider $n$ of
them and not the infinite number required to expand the whole Hilbert
space} $\mathbb{I}_{a}$ and $\mathbb{I}_{b}$ before Eq (24) of \citep{Avez2012}
is not justified as they are only ``identities'' over the subspaces
spanned by the corresponding bases. In addition, the removal of non-occupied
states described in their appendix B and required to obtain their
Eqs (61) and (54) assumes unitary overlap matrices in contradiction
with the use of pseudo-identities in their derivation that necessarily
introduce non-unitary overlap matrix. Although the final result is
correct, the derivation misses important aspects of the problem. 

It is also worth to mention that the result of Eqs (57-59) in \citep{PhysRevC.84.014307},
dealing with non-equivalent bases but using an orthogonal version
of the union of the two basis is fully equivalent to Eq \eqref{eq:PfaffFin}
above as demonstrated in the appendix.

Finally, in Ref \citep{Scamps2013} a formula is given to compute
the overlap between two BCS wave functions, where the canonical states
of the BCS transformation are not equivalent under unitary transformations.
The formula is based on the Pfaffian formalism and is taken directly
from Eq (5) in Ref \citep{Bertsch2012}. The formula given in \citep{Bertsch2012}
was obtained implicitly assuming that the basis is complete under
the action of the symmetry operator introduced in that paper and therefore
the derivation is not paying attention to the problems associated
with non-equivalent bases. Therefore it seems surprising at a first
sight that the formula given in \citep{Scamps2013} is giving the
correct result of Eq \eqref{eq:PfaffFin} (as demonstrated in the
appendix where the connection between Eq (7) of \citep{Bertsch2012}
and this equation is given). Apart from the not-so-simple derivation
given in the appendix, there is a simple argument to support this
surprising coincidence. As in \citep{Scamps2013} let us consider
two BCS wave functions $|\phi_{a}\rangle=\prod_{k}(u_{k}^{a}+v_{k}^{a}a_{k}^{+}a_{\bar{k}}^{+})|-\rangle$
given in basis $B_{A}=\{a_{k}^{+},k=1,\ldots,N_{a}\}$ and $|\phi_{b}\rangle=\prod_{l}(u_{l}^{b}+v_{l}^{b}b_{l}^{+}b_{\bar{l}}^{+})|-\rangle$
given in basis $B_{B}=\{b_{l}^{+},l=1,\ldots,N_{b}\}$. One can expand
$b_{l}^{+}=\sum_{k}R_{kl}a_{k}^{+}$ where the sum in $k$ extends
to all elements in a complete basis of the $a_{k}^{+}$ (i.e. $N_{a}\rightarrow\infty)$.
One can write $b_{l}^{+}=b_{l}^{(0)\,+}+b_{l}^{(1)\,+}$ where $b_{l}^{(0)\,+}$
corresponds to the terms in the sum with $k$ in the finite basis
$B_{A}$ (i.e. $N_{a}$ finite) and $b_{l}^{(1)\,+}$ to the remaining
terms in the sum. The key argument is that $\{b_{l}^{(1)\,+},a_{k}\}=0$
with $k$ in $B_{A}$ what allows to move the $b_{l}^{(1)\,+}$ in
the product\begin{widetext}
\[
\langle\phi_{a}|\phi_{b}\rangle=\prod_{kl}\langle-|(u_{k}^{a}+v_{k}^{a}a_{\bar{k}}a_{k})(u_{l}^{b}+v_{l}^{b}(b_{l}^{(0)\,+}+b_{l}^{(1)\,+})(b_{\bar{l}}^{(0)\,+}+b_{\bar{l}}^{(1)\,+})|-\rangle
\]
\end{widetext} to the left in order to be annihilated by the vacuum.
As a consequence
\[
\langle\phi_{a}|\phi_{b}\rangle=\prod_{kl}\langle-|(u_{k}^{a}+v_{k}^{a}a_{\bar{k}}a_{k})(u_{l}^{b}+v_{l}^{b}b_{l}^{(0)\,+}b_{\bar{l}}^{(0)\,+})|-\rangle
\]
which leads to the expected result. It is important to note that,
although the argument is valid for the overlap, is does not apply
in the evaluation of overlaps $\langle\phi_{a}|\hat{O}|\phi_{b}\rangle$
of general operators, like the one body operator $\hat{O}=\sum_{kl}O_{kl}^{A}a_{k}^{+}a_{l}=\sum_{kl}O_{kl}^{B}b_{k}^{+}b_{l}=\sum_{kl}O_{kl}^{AB}a_{k}^{+}b_{l}$
(defined with an obvious notation) because now, due to the presence
of $\hat{O}$ in the middle of the product the $b_{l}^{(1)\:+}$operators
can not freely jump over $\hat{O}$ to its left in order to be annihilated
by the left vacuum.

\section{Conclusions}

In this paper I discuss how to handle the calculation of overlaps
between HFB wave functions using the pfaffian formalism in the common
situation where they are expressed in non-equivalent single particle
basis. The result obtained expanding the bases to cover the whole
Hilbert space is proven to be equivalent to previous result \citep{PhysRevC.84.014307}
using an orthogonalized version of the union of the two basis. Comparison
with the work of Scamps et al \citep{Scamps2013} clarifies the reason
why their pfaffian formula works in this specific case in spite of
not considering at all the issue with non-equivalent basis in their
developments. Finally, some inconsistencies in other derivation by
Avez et al \citep{Avez2012} are pointed out and discussed.
\begin{acknowledgments}
This work has been supported by the Spanish Ministerio de Ciencia,
Innovación y Universidades and the European regional development fund
(FEDER), grants No PID2021-127890NB-I00.
\end{acknowledgments}

\appendix

\section{Relating pfaffian formulas}

In this appendix it is proven that Eq (59) of \citep{PhysRevC.84.014307}
and Eqs (61) and (54) of \citep{Avez2012} are the same in spite of
being obtained in a rather different manner. In \citep{PhysRevC.84.014307}
the basis $U_{0}\cup U_{1}$ union of the two bases $B_{0}$ and $B_{1}$
was used to deal with non-equivalent bases. This approach requires
to consider the orthogonalization of $U_{0}\cup U_{1}$ by diagonalization
of the norm matrix
\[
\mathcal{N}=\left(\begin{array}{cc}
\mathbb{I} & T\\
T^{+} & \mathbb{I}
\end{array}\right)
\]
where $T_{ij}=_{0}\langle i|j\rangle_{1}$ is the overlap matrix.
This matrix is not required to be a square matrix (i.e. $B_{0}$ and
$B_{1}$ might have different dimensions) but we will assume it to
be so for simplicity. The singular value decompostion of $T=E\Delta F^{+}$
with $E$ and $F$ unitary and $\Delta$ diagonal and positive definite
allows to write the norm matrix as
\[
\mathcal{N}=\left(\begin{array}{cc}
E & 0\\
0 & F
\end{array}\right)\left(\begin{array}{cc}
\mathbb{I} & \Delta\\
\Delta & \mathbb{I}
\end{array}\right)\left(\begin{array}{cc}
E^{+} & 0\\
0 & F^{+}
\end{array}\right)=D\left(\begin{array}{cc}
\mathbb{I} & \Delta\\
\Delta & \mathbb{I}
\end{array}\right)D^{+}
\]
 and its inverse
\begin{equation}
\mathcal{N}^{-1}=D\left(\begin{array}{cc}
\left(\mathbb{I}-\Delta^{2}\right)^{-1} & -\Delta\left(\mathbb{I}-\Delta^{2}\right)^{-1}\\
-\Delta\left(\mathbb{I}-\Delta^{2}\right)^{-1} & \left(\mathbb{I}-\Delta^{2}\right)^{-1}
\end{array}\right)D^{+}\label{eq:InvN}
\end{equation}
Please note that the four blocks of 
\[
\bar{S}=\left(\begin{array}{cc}
\left(\mathbb{I}-\Delta^{2}\right)^{-1} & -\Delta\left(\mathbb{I}-\Delta^{2}\right)^{-1}\\
-\Delta\left(\mathbb{I}-\Delta^{2}\right)^{-1} & \left(\mathbb{I}-\Delta^{2}\right)^{-1}
\end{array}\right)
\]
are diagonal matrices. The overlap in Eq (59) of \citep{PhysRevC.84.014307}
is given by
\[
\langle\phi_{0}|\phi_{1}\rangle=s_{2N}\textrm{pf}\tilde{\mathbb{M}}
\]
where
\[
\tilde{\mathbb{M}}=\left(\begin{array}{cc}
N^{(1)} & -\mathbb{I}\\
\mathbb{I} & -N^{(0)\,*}
\end{array}\right)
\]
with $\tilde{N}^{(i)}=\left(\mathcal{N}^{1/2}\right)^{+}\tilde{M}_{E}^{(i)}\left(\mathcal{N}^{1/2}\right)^{*}$
and
\[
\tilde{M}_{E}^{(1)}=\left(\begin{array}{cc}
0 & 0\\
0 & M^{(1)}
\end{array}\right),\;\tilde{M}_{E}^{(0)}=\left(\begin{array}{cc}
M^{(0)} & 0\\
0 & 0
\end{array}\right).
\]
With these definitions one can write
\[
\langle\phi_{0}|\phi_{1}\rangle=s_{2N}\det\mathcal{N}\textrm{pf}\left(\begin{array}{cc}
\tilde{M}_{E}^{(1)} & -\mathcal{N}^{-1}\\
\left(\mathcal{N}^{*}\right)^{-1} & -\tilde{M}_{E}^{(0)\,*}
\end{array}\right)
\]
and using now Eq \eqref{eq:InvN} for the inverse of $\mathcal{N}$one
obtains

\begin{equation}
\langle\phi_{0}|\phi_{1}\rangle=s_{2N}\det\mathcal{N}\textrm{pf}\left(\begin{array}{cc}
\bar{M}_{E}^{(1)} & -\bar{S}\\
\bar{S}^{*} & -\bar{M}_{E}^{(0)\,*}
\end{array}\right)\label{eq:overC}
\end{equation}
with 
\[
\bar{M}_{E}^{(1)}=\left(\begin{array}{cc}
0 & 0\\
0 & F^{+}M^{(1)}F^{*}
\end{array}\right),\;\bar{M}_{E}^{(0)}=\left(\begin{array}{cc}
E^{+}M^{(0)}E & 0\\
0 & 0
\end{array}\right).
\]
The argument of the pfaffian in Eq \eqref{eq:overC} acquires the
$4\times4$ block structure\textbackslash
\[
\mathbb{M}=\begin{pmatrix}\begin{array}{cc|cc}
0 & 0 & -\bar{S}_{11} & -\bar{S}_{12}\\
0 & F^{+}M^{(1)}F^{*} & -\bar{S}_{12} & -\bar{S}_{22}\\
\hline \bar{S}_{11} & \bar{S}_{12} & -\left(E^{+}M^{(0)}E\right)^{*} & 0\\
\bar{S}_{12} & \bar{S}_{22} & 0 & 0
\end{array}\end{pmatrix}
\]
that can be transformed to the form 
\[
\mathbb{M}'=\begin{pmatrix}\begin{array}{cc|cc}
-\left(E^{+}M^{(0)}E\right)^{*} & \bar{S}_{12} & \bar{S}_{11} & 0\\
-\bar{S}_{12} & F^{+}M^{(1)}F^{*} & 0 & -\bar{S}_{11}\\
\hline -\bar{S}_{11} & 0 & 0 & -\bar{S}_{12}\\
0 & \bar{S}_{11} & \bar{S}_{12} & 0
\end{array}\end{pmatrix}
\]
by means of a congruence transformation (see appendix A of \citep{PhysRevC.84.014307})
with determinant $(-1)^{N_{}}$. Applying now the formula for the
pfaffian of a bipartite matrix (see appendix B of \citep{PhysRevC.84.014307})
\[
\textrm{pf\ensuremath{\mathbb{M}}'=}\textrm{pf}\begin{pmatrix}0 & -\bar{S}_{12}\\
\bar{S}_{12} & 0
\end{pmatrix}\textrm{pf}\begin{pmatrix}-\left(E^{+}M^{(0)}E\right)^{*} & \Delta^{-1}\\
-\Delta^{-1} & F^{+}M^{(1)}F^{*}
\end{pmatrix}
\]
because $\Delta^{-1}=\bar{S}_{12}-\bar{S}_{11}\bar{S}_{12}^{-1}\bar{S}_{11}$.
Using now a few pfaffian properties one arrives to
\[
\textrm{pf}\ensuremath{\mathbb{M}}'=(-)^{N}\det\bar{S}_{12}\det E\det F^{*}\textrm{pf}\begin{pmatrix}M^{(1)} & T^{-1}\\
-T^{-1} & -M^{(0)\,*}
\end{pmatrix}
\]
Taking into account that 
\[
\det T=\det E\det F^{*}\prod_{i}\Delta_{i},
\]
$\det\mathcal{N}=\prod_{i}(1-\Delta_{i}^{2})$ and $\det\bar{S}_{12}=(-1)^{N}\prod_{i}\Delta_{i}/(1-\Delta_{i}^{2})$
it is easy to obtain
\[
\langle\phi_{0}|\phi_{1}\rangle=(-1)^{n}\det T\textrm{pf}\begin{pmatrix}M^{(1)} & T^{-1}\\
-T^{-1} & -M^{(0)\,*}
\end{pmatrix}
\]
which is the result of Aver et al \citep{Avez2012}.

In the supplemental material of \citep{Bertsch2012} it is demonstrated
how the main result of \citep{PhysRevC.79.021302} given in its Eq
(7) is equivalent to Eq (7) of \citep{Bertsch2012} for the special
case $\mathcal{R}=\mathbb{I}$ . The equivalence of the two formulas
is proven by using congruence transformations on the affected matrices
and the properties of the pfaffian under such kind of transformations.
One can use the same kind of arguments to prove that the pfaffian
in Eq \ref{eq:PfaffFin} above is related to the pfaffian of Eq (7)
of \citep{Bertsch2012} involving the matrix
\[
\mathbb{M}=\left(\begin{array}{cc}
V_{0}^{T}U_{0} & V_{0}^{T}\mathcal{R}V{}_{1}^{*}\\
-V_{1}^{+}\mathcal{R}^{T}V_{0} & U_{1}{}^{+}V_{1}^{*}
\end{array}\right)
\]
It is straightforward to prove that the above matrix can be block
diagonalized by means of a congruence
\[
\mathbb{M}=X^{T}\left(\begin{array}{cc}
\left(\bar{M}^{(0)\:*}\right)^{-1} & 0\\
0 & -\left(\bar{M}^{(1)}\right)^{-1}+\mathcal{R}^{T}\bar{M}^{(0)\:*}\mathcal{R}
\end{array}\right)X
\]
with 
\[
X=\left(\begin{array}{cc}
V_{0} & \bar{M}^{(0)\,*}\mathcal{R}V_{1}^{*}\\
0 & V_{1}^{*}
\end{array}\right)
\]
As a consequence,
\begin{align*}
\textrm{pf}\mathbb{M} & =\det V_{0}\det V_{1}^{*}\textrm{pf}\left(\bar{M}^{(0)\:*}\right)^{-1}\textrm{\ensuremath{\times}}\\
 & \times\textrm{pf}\left[-\left(\bar{M}^{(1)}\right)^{-1}+\mathcal{R}^{T}\bar{M}^{(0)\:*}\mathcal{R}\right]
\end{align*}
In the same way 
\[
\mathbb{M}'=\left(\begin{array}{cc}
\mathcal{R}\bar{M}^{(1)}\mathcal{R}^{T} & -\mathbb{I}\\
\mathbb{I} & -\bar{M}^{(0)\,*}
\end{array}\right)
\]
can also be block diagonalized 
\[
\mathbb{M}'=Y^{T}\left(\begin{array}{cc}
\bar{M}^{(1)} & 0\\
0 & -\bar{M}^{(0)\:*}+\left(\mathcal{R}\bar{M}^{(1)}\mathcal{R}^{T}\right)^{-1}
\end{array}\right)Y
\]
with 
\[
Y=\left(\begin{array}{cc}
\mathcal{R}^{T} & -\left(\mathcal{R}M^{(1)}\right)^{-1}\\
0 & \mathbb{I}
\end{array}\right)
\]
and 
\begin{align*}
\textrm{pf}\mathbb{M}' & =\det\mathcal{R}\textrm{pf}\bar{M}^{(1)}\textrm{pf \ensuremath{\left[-\bar{M}^{(0)\,*}+\left(\mathcal{R}\bar{M}^{(1)}\mathcal{R}^{T}\right)^{-1}\right]}}\\
 & =\textrm{pf}\bar{M}^{(1)}\textrm{pf \ensuremath{\left[-\mathcal{R}^{T}\bar{M}^{(0)\,*}\mathcal{R}+\left(\bar{M}^{(1)}\right)^{-1}\right]}}.
\end{align*}
To reduce further the expressions we need the properties $\textrm{pf}A=(-1)^{n}/\textrm{pf}A$
and $\textrm{pf}(-A)=(-1)^{n}\textrm{pf}A$ valid for matrices of
dimension $M=2n$. Using them one obtains 
\[
\textrm{pf\ensuremath{\mathbb{M}}=}\frac{\det V_{0}}{\textrm{pf\ensuremath{\bar{M}^{(0)\:*}}}}\frac{\det V_{1}^{*}}{\textrm{pf\ensuremath{\bar{M}^{(1)}}}}\textrm{pf}\ensuremath{\mathbb{M}'}
\]
that is the desired result as $\det V/\textrm{pf}M^{*}=\textrm{pf}(U^{T}V)$
which is the normalization factor connecting the wave functions defined
in \citep{Bertsch2012} and in \citep{PhysRevC.79.021302}. The derivation
has been carried out without any assumption on the properties of $\mathcal{R}$and
therefore Eq (7) of \citep{Bertsch2012} is also valid in the case
of non unitary $\mathcal{R}$ corresponding to non-equivalent basis.

\bibliographystyle{apsrev4-2}
%

\end{document}